\documentclass[a4paper]{article}
\usepackage[a4paper]{geometry}
\usepackage{amsfonts,amssymb,amsmath}
\title{On the Esscher transforms and other equivalent martingale measures
for Barndorff-Nielsen and Shephard stochastic volatility models with
jumps\thanks{Financial support from the
Austrian Science Fund (FWF) under grant P15889 and grant 2006132713--005
of the Italian Ministry for Scientific Research
is gratefully acknowledged.}}
\author{Friedrich Hubalek\thanks{Corresponding author.
Tel.: +43 1 58801 10513; fax: +43 1 58801 10599},\\
{\small Vienna University of Technology,
Financial and Actuarial Mathematics,} \\
{\small Wiedner Hauptstra\ss{}e~8/105--1,
A--1040 Vienna, Austria (fhubalek@fam.tuwien.ac.at)}
\and
Carlo Sgarra,\\
{\small Department of Mathematics,
Politecnico di Milano,}\\
{\small Piazza Leonardo da Vinci, 32,
I--20133 Milan, Italy (carlo.sgarra@polimi.it)}}
\date{\tt\jobname.tex(\today)}
\def\Rheinlaender{Rheinl\"ander}

\def\Levy{L\'evy}
\def\Ito{It\^{o}}

\def\qed{\mbox{}\hfill$\Box$}
\def\textfrac#1#2{{\textstyle\frac{#1}{#2}}}
\newtheorem{theorem}{Theorem}
\newtheorem{remark}{Remark}
\newtheorem{lemma}{Lemma}

\begin{document}
\maketitle
\begin{abstract}
We compute resp.\ discuss the Esscher martingale transform for exponential
processes, the Esscher martingale transform for linear processes, the minimal
martingale measure, the class of structure preserving martingale measures,
and the minimum entropy martingale measure for stochastic volatility models of
Ornstein-Uhlenbeck type as introduced by Barndorff-Nielsen and Shephard.
We show, that in the model with leverage, with jumps both in the volatility and in the
returns, all those measures are different, whereas in the model without
leverage, with jumps in the volatility only and a continuous return process,
several measures coincide, some simplifications can be made and the results are more explicit.
We illustrate our results with parametric examples used in the
literature.
\end{abstract}
\section{Introduction}
\Levy{} processes provide a lot of flexibility in financial modelling.
Although financial returns increments exhibit some kind of serial
dependence, many of their essential features are captured by this class of
models: heavy tails, aggregational Gaussianity, volatility clustering are
some of their features easily described by means of models based on \Levy{}
processes. Introduction of jumps anyway rises the problem of dealing with
incomplete market models; that means that there exist infinitely many
martingale measures, compatible with the no arbitrage requirement and
equivalent to the physical measure describing the underlying evolution, one
can use to price derivative securities.

One reasonable way to solve this
problem is based on the observation that in incomplete markets the "correct"
equivalent martingale measure (EMM from now on) could not be independent on
the preferences of investors any more, so by guessing a suitable utility
function describing these preferences, the "optimal" EMM should maximize the
expected value of this utility. It has been proved that for many interesting
cases of utility functions this problem admits a dual formulation: finding
an EMM\ maximizing some classes of utility functions is equivalent to find
EMM minimizing some kind of distances \cite{FF}. Of particular relevance in
the framework of utility maximization are the equivalent martingale
measures maximizing exponential utility (and minimizing, by duality, the
relative entropy) and those maximizing quadratic utility (and minimizing,
always by duality, an appropriate $L^2$-distance).

Another popular approach to option
pricing for incomplete models had been related to the construction of the
Esscher martingale transform. As it has been already pointed out in
\cite{KS} two different Esscher martingale transforms exist
for \Levy{} processes
according to the choice of the parameter which defines the measure: one
turning the ordinary exponential process into a martingale and another one
turning into a martingale the stochastic exponential. They have been called
the Esscher martingale transform for the exponential process and the Esscher
martingale transform for the linear process respectively. It has been shown
in  \cite{ES} that for exponential \Levy{} models the Esscher martingale
transform for the linear process is also the minimal entropy martingale measure,
i.e., the equivalent martingale measure which minimizes the relative entropy,
and that this measure has also the property of preserving the \Levy{}
structure of the model, see also \cite{levyrep}.

The definition and the abstract theory of the
Esscher martingale transforms for general semimartingales has been given in
\cite{KS}, following previous results in discrete
time in~\cite{BDES1996} and~\cite{BDES1998}.
Some recent results related to generalization of the Esscher transform to a non-\Levy{} setting
are in \cite{BM07,STY4,ECS}.

Since a few years interest is grown also in a "second generation" of models
based on \Levy{} processes, i.e., the stochastic volatility models driven by
\Levy{} processes; the model introduced by O.E.~Barndorff-Nielsen
and N.~Shephard belongs to this class \cite{BNS1}, \cite{BNS2}.
In this model, or we
could better say, class of models, the volatility is described by an
Ornstein-Uhlenbeck process driven by a \Levy{} process with positive
increments, i.e., a subordinator.

For these models (from now on BNS) some results are already
available in the context of option pricing:
E.~Nicolato and E.~Venardos \cite{NV} introduced a class of
{\em structure preserving} equivalent martingale
measures, under which the stochastic process describing the evolution of the
underlying asset follows a stochastic differential equation with the same
structure although with possibly different parameters. Under such measures
the problem of pricing options could be solved by using a transform-based
technique.

The  purpose of this paper is to give  the explicit construction of the
Esscher transform both for the linear and for the exponential processes and
the minimal martingale measure for BNS models and to present a systematic
comparison of all the measures available in the literature (including those
we will obtain in the present work) and their relations.

In Section~\ref{Sec-BNS} we will recall the essential features of
BNS models and we will
give the explicit calculations of the characteristic triplet characterizing
them as a semimartingale process, in order to apply the general theory of
the Esscher transform introduced in \cite{KS} in Section~\ref{Sec-Esscher}.
There we will present the explicit construction of the Esscher martingale transforms both
for the linear and the exponential processes.

In Section~\ref{Sec-Examples} we will discuss the existence of the
Esscher martingale transforms for some relevant examples of BNS models,
and present their construction, if they exist.

In Section~\ref{sec-other} we will recall the
main results obtained in \cite{NV} about the equivalent martingale measures
which preserves the model structure and we will show that they do not
coincide with neither of the Esscher martingale transforms.
Next we will give the expression of the
minimal martingale measure for BNS.
At the end of this section we briefly recall the main result obtained for
the minimal entropy martingale measure in the
no-leverage case, and compare it with the previously discussed measures.

Finally in Section~\ref{simplifications} we show that both Esscher transforms and
the minimal martingale measure coincide in the non-leverage case.
We then recall the results of \cite{EBMB} for the minimal entropy martingale measure in the
non-leverage case.

In appendices~\ref{app-a}--\ref{app-c} we give sufficient conditions to
assure that the candidates for the density processes for the exponential
Esscher, the linear Esscher and the minimal martingale measure
exist and are proper martingales.

We use the BNS model, because it is a model that exhibits a connection
of jumps and stochastic volatility, but yet allows very explicit calculations.
In \cite{CT} it is argued, that the Bates model is simpler and sufficient
for most purposes (and thus perhaps preferable), but as jumps and
stochastic volatility are independent in that model, it is less
interesting from a mathematical perspective.

Throughout the paper we use the notation of \cite{JS} for semimartingale
theory, stochastic calculus, and stochastic integration.
In particular, if $X$ is a semimartingale, then $L(X)$ denotes
the set of predictable $X$-integrable processes, and for $H\in L(X)$
the stochastic integral is sometimes written as $H\cdot X$.
\section{The BNS model}\label{Sec-BNS}
\subsection{Specification of the model}
We will now focus our attention on the class of stochastic volatility models
with jumps, that has been introduced  by Barndorff-Nielsen and Shephard
in \cite{BNS1,BNS2}.

Suppose we are given
a probability space $(\Omega,\mathcal F,P)$ carrying a standard
Brownian motion $W$ and an independent increasing pure jump \Levy{} process $Z$.
The process~$Z$ is called the {\em background driving \Levy{} process}, or BDLP
for short.
We assume that the discounted stock price is given by
\begin{equation}\label{SX}
S_t=S_0e^{X_t},
\end{equation}
where $S_0>0$ is a constant, logarithmic returns satisfy
\begin{equation}\label{SDE-X}
dX_t=(\mu+\beta V_{t-})dt+\sqrt{V_{t-}}dW_t+\rho dZ_{\lambda t},
\end{equation}
starting from $X_0=0$,
and the instantaneous variance satisfies
\begin{equation}\label{dV}
dV_t=-\lambda V_{t-}dt+dZ_{\lambda t}
\end{equation}
with constant initial value $V_0>0$.
The parameter range is
\begin{equation}
\mu\in\mathbb R,\quad
\beta\in\mathbb R,\quad
\rho\leq0,\quad
\lambda>0.
\end{equation}
We denote the cumulant function and the \Levy{} measure of $Z$ by $k(z)$ resp.\ $U(dx)$.
Since $Z$ is increasing we have
\begin{equation}
k(z)=\int_0^{\infty}(e^{zx}-1)U(dx).
\end{equation}
We will work with the {\em usual natural} filtration $(\mathcal F_t)$ generated by the
pair $(W_t,Z_{\lambda t})$, as it is defined in \cite[2.63, p.63]{HWY}.
The solution to~(\ref{dV}) is
\begin{equation}
V_t=V_0e^{-\lambda t}+\int_0^te^{-\lambda(t-s)}dZ_{\lambda s}.
\end{equation}
Therefore we have the inequalities
\begin{equation}
V_0e^{-\lambda t}\leq V_t\leq V_0e^{-\lambda t}+Z_{\lambda t}.
\end{equation}
The evolution of the stock price is governed by
\begin{equation}
dS_t=S_{t-}d\tilde X_t,
\end{equation}
with
\begin{equation}
\tilde X_t=\int_0^tS_{u-}^{-1}dS_u.
\end{equation}
We will discuss the process $\tilde X$, which is called
the exponential transform of $X$, in more detail below.
\begin{remark}
In principle the leverage parameter $\rho$ could be an arbitrary real number.
If $\rho=0$ we call the model a BNS model without leverage.
In that case the trajectories of logarithmic returns, and thus of
the asset price are continuous.
If $\rho\neq0$ we call the model a BNS model with leverage,
and returns and the asset price exhibit jumps.
If $\rho>0$ and the jumps of $Z$ are unbounded, the
asset price process is not locally bounded.
If $\rho\leq0$ the asset price process will be locally bounded.
Typically $\rho\leq0$ and we restrict our analysis to that case.
\end{remark}
\begin{remark}
As $V$ is of finite variation and $W$ is continuous we could define
the model pathwise by Riemann-Stieltjes integrals without reference
to stochastic integration.
\end{remark}
\subsection{Semimartingale characteristics and cumulants for logarithmic returns}
The background driving \Levy{} process~$Z$, or BDLP for short,
is increasing, and as a consequence we can always use the {\em zero truncation
function} $h(x)=0$, and we shall do so for most of the paper.
If we assume also that $E[Z_1]<\infty$, we can use the
{\em identity truncation function} $h(x)=x$, which corresponds
to the Doob-Meyer decomposition of a special semimartingale.

For later usage it is convenient to introduce
\begin{equation}
U_\rho(dx)=U\left(\frac{dx}{\rho}\right),
\end{equation}
and
\begin{equation}
k_\rho(z)=k(\rho z),
\end{equation}
which are simply the \Levy{} measure resp.\ the cumulant function
corresponding to the process~$\rho Z$.
Let us recall that the jumps of the processes $X$ and $Z$ are related
by $\Delta X_t=\rho\Delta Z_{\lambda t}$
and therefore the jump measure~$\mu_X(dx,dt)$
of the process~$X$ is related to the jump measure~$\mu_Z(dx,dt)$
of the process~$Z$ by
\begin{equation}\label{muXZ}
\mu_X(dx,dt)=\mu_Z\left(\frac{dx}{\rho},\lambda dt\right).
\end{equation}
We will denote by $\nu(dx,dt)$ the predictable compensator of $\mu_X(dx,dt)$.
Let us provide now the semimartingale characteristics of $X$.
\begin{lemma}\label{char-X}
The semimartingale characteristics of $X$
with respect to the zero truncation function
are given by $(B,C,\nu)$, which satisfy
\begin{equation}
dB_t=b_tdt,\quad
dC_t=c_tdt,\quad
\nu(dt,dx)=F(t,dx)dt,
\end{equation}
where
\begin{equation}
b_t=\mu+\beta V_{t-},\quad
c_t=V_{t-},\quad
F(t,dx)=\lambda U_\rho(dx).
\end{equation}
\end{lemma}
Proof: The jump measure of $(X_t)$ coincides trivially with the jump measure of the \Levy{}
process~$\rho Z_{\lambda t}$. Hence its predictable compensator is the measure $\lambda U_\rho(dx)dt$.
Using the definition of semimartingale characteristics as given in \cite[II.2a, p.75f]{JS}
for $h(x)=0$ we have
\begin{equation}
X_t-\sum_{s\leq t}\Delta X_s=
\int_0^t(\mu+\beta V_{s-})ds+\int_0^t\sqrt{V_{s-}}dW_s,
\end{equation}
and we can identify the corresponding drift and the quadratic variation of
the continuous martingale part.\hfill~\qed
\begin{remark}
If we used the identity truncation function, we had $\mbox{b}_t=
\mu+\rho\lambda\zeta+\beta V_{t-}$ with $\zeta=E[Z_1]$
for the first differential characteristic.
\end{remark}
For the next lemma we need the notions of an exponentially special
semimartingale and its (modified) Laplace cumulant process
from~\cite[Def.2.12, p.402 and Def.2.16, p.403]{KS}.
\begin{lemma}
Let $\theta\in L(X)$ be such that $\theta\cdot X$ is exponentially special.
The modified Laplace cumulant process of $X$ in $\theta$ is then given by
\begin{equation}
K^X(\theta)_t=\int_0^t\tilde\kappa^X(\theta)_sds,
\end{equation}
where
\begin{equation}
\tilde\kappa^X(\theta)_t=b_t\theta_t+\frac12c_t\theta_t^2+\lambda k(\rho\theta_t).
\end{equation}
\end{lemma}
Proof: This follows immediately from the characteristics computed above
and \cite[Theorem~2.18.1, p.404]{KS}.\hfill~\qed
\subsection{Semimartingale characteristics and cumulants for the exponential
transform}
In the following it is useful to rewrite the ordinary exponential in~(\ref{SX})
as stochastic exponential. This can be done
by using $\tilde X$, the exponential transform of $X$.
and we obtain
\begin{equation}
S_t=S_0\mathcal E(\tilde X)_t.
\end{equation}
According to \cite[Lemma 2.6.1, p.399]{KS} we have
$\tilde X=X+\frac12\langle X^c,X^c\rangle+(e^x-1-x)\ast\mu_X$.
This means
\begin{equation}
\begin{split}
\tilde X_t
=
&\int_0^t(\mu+\beta V_{s-})ds
+\int_0^t\sqrt{V_{s-}}dW_s
+\rho Z_{\lambda t}
\\
&+
\frac12\int_0^tV_{s-}ds
+\sum_{s\leq t}(e^{\rho\Delta Z_{\lambda s}}-1-\rho\Delta Z_{\lambda s})
\end{split}
\end{equation}
This can be rewritten as
\begin{equation}
\tilde X_t=
\int_0^t(\mu+\tilde\beta V_{s-})ds+
\int_0^t\sqrt{V_{s-}}dW_s+
\sum_{s\leq t}(e^{\rho\Delta Z_{\lambda s}}-1),
\end{equation}
where
\begin{equation}
\tilde\beta=\beta+\frac12.
\end{equation}
We have clearly
\begin{equation}
\tilde X^c=X^c
\end{equation}
and the jumps are
\begin{equation}
\Delta\tilde X_t=
e^{\rho\Delta Z_{\lambda t}}-1.
\end{equation}
We want to introduce the process
\begin{equation}
M_t=\rho Z_{\lambda t},
\end{equation}
and its exponential transform given by
\begin{equation}\label{tildeM}
\tilde M_t=\sum_{s\leq t}(e^{\rho\Delta Z_{\lambda s}}-1).
\end{equation}
Then we can write
\begin{equation}
dX_t=
(\mu+\beta V_{t-})dt+\sqrt{V_{t-}}dW_t+dM_t,
\end{equation}
and
\begin{equation}
d\tilde X_t=(\mu+\tilde\beta V_{t-})dt+\sqrt{V_{t-}}dW_t+d\tilde M_t.
\end{equation}
We note that $\tilde M$ is a \Levy{} process.
It is helpful to introduce the function
\begin{equation}
g_\rho(x)=e^{\rho x}-1,
\end{equation}
its inverse function
\begin{equation}
g_\rho^{-1}(x)=\frac1\rho\ln(1+x),
\end{equation}
and the induced measure
\begin{equation}
\tilde U_\rho=U\circ g_\rho^{-1}.
\end{equation}
If $U$ admits a density $u$ then $\tilde U_\rho$ admits a density
$\tilde u_\rho$ given by
\begin{equation}
\tilde u_\rho(x)=
\frac1{\rho(1+x)}u\left(
\frac1\rho\ln(1+x)
\right).
\end{equation}
The cumulant function of $\tilde M$ is
\begin{equation}\label{ktilderho}
\tilde k_\rho(z)=\int_0^\infty(e^{z(e^{\rho x}-1)}-1)U(dx).
\end{equation}
\begin{lemma}
The semimartingale characteristics of $\tilde X$
with respect to the zero truncation function
are given
by $(\tilde B,\tilde C,\tilde\nu)$, which satisfy
\begin{equation}
d\tilde B_t=\tilde b_tdt,\quad
d\tilde C_t=\tilde c_tdt,\quad
\tilde\nu(dt,dx)=
\tilde F(t,dx)dt,
\end{equation}
where
\begin{equation}
\tilde b_t=\mu+\tilde\beta V_{t-}\quad
\tilde c_t=V_{t-}\quad
\tilde F(t,dx)=\lambda\tilde U_\rho(dx).
\end{equation}
\end{lemma}
Proof: This follows from the characteristics of $X$ given above and
\cite[Theorem~II.8.10, p.136]{JS}.\hfill~\qed.
\begin{remark}
If we used the identity truncation function, we had $\tilde b_t=
\mu+\lambda k(\rho)+\beta V_{t-}$
for the first differential characteristic.
\end{remark}
In the following lemma we need the notion of
the derivative of a cumulant process from \cite[Def.2.22, p.407]{KS}.
\begin{lemma}
Let $\theta\in L(\tilde X)$ be such that $\theta\cdot\tilde X$
is exponentially special.
The modified Laplace cumulant process of $\tilde X$ in
$\theta$ is then given by
\begin{equation}
K^{\tilde X}(\theta)_t=\int_0^t\tilde\kappa^{\tilde X}(\theta)_sds,
\end{equation}
where
\begin{equation}
\tilde\kappa^{\tilde X}(\theta)_t=
\tilde b_t\theta_t+\frac12\tilde c_t\theta_t^2+\lambda\tilde k_\rho(\theta_t).
\end{equation}
The derivative of the cumulant process $K^{\tilde X}(\theta)$
is given by
\begin{equation}
DK^{\tilde X}(\theta)=
\int_0^t\tilde\kappa^{\tilde X}(\theta)_sds,
\end{equation}
where
\begin{equation}
D\tilde\kappa^{\tilde X}(\theta)_t=
\tilde b_t+\tilde c_t\theta_t+\lambda\tilde k_\rho'(\theta_t).
\end{equation}
\end{lemma}
Proof: The expression for $K^{\tilde X}(\theta)$ follows immediately from the
characteristics of $\tilde X$ computed above
and \cite[Theorem~2.18.1--2, p.404]{KS}.
The expression for $DK^{\tilde X}(\theta)$ follows
from \cite[Definition~2.22, p.407]{KS}.\hfill~\qed
\section{Esscher martingale transforms for BNS models}\label{Sec-Esscher}
\subsection{The Esscher martingale transform for exponential processes}
Let us look at the Esscher martingale transform for exponential processes
as described in \cite[Theorem~4.1, p.421]{KS}.
The discounted asset price $S$ satisfies
$S=S_0e^X$.
We have to find the solution to
\begin{equation}\label{KX1=0}
K^X(\theta+1)-K^X(\theta)=0.
\end{equation}
Suppose now we can establish a solution $\theta^\sharp_t$
to that equation, almost surely for all $t\in[0,T]$, and
$G^\sharp_t=e^{\theta^\sharp\cdot X_t-K^X(\theta^\sharp)_t}$
defines a martingale $(G^\sharp_t)_{0\leq t\leq T}$, then
we can define a probability measure $P^\sharp$ by
\begin{equation}
\frac{dP^\sharp}{dP}=
e^{\theta^\sharp\cdot X_T-K^X(\theta^\sharp)_T}.
\end{equation}
This measure is then the Esscher martingale transform for the exponential
process~$e^X$. If there is no solution with the required properties,
we say the Esscher martingale transform for the exponential process
does not exist.
\begin{theorem}\label{thm-sharp}
Suppose there is $\theta^\sharp\in L(X)$, such that
$\theta^\sharp\cdot X$ is exponentially special,
\begin{equation}\label{eq-sharp}
K^X(\theta^\sharp+1)-K^X(\theta^\sharp)=0,
\end{equation}
and
\begin{equation}\label{Gsharp}
G^\sharp_t=\mathcal E(\tilde N^\sharp)_t,
\end{equation}
with
\begin{equation}
\tilde N^\sharp_t=\int_0^t\psi^\sharp_sdW_s+\int_0^t\int(Y^\sharp(s,x)-1)(\mu_X-\nu)(dx,ds),
\end{equation}
\begin{equation}
\psi^\sharp_t=\theta^\sharp_t\sqrt{V_{t-}}
\end{equation}
and
\begin{equation}
Y^\sharp(t,x)=e^{\theta^\sharp_t\rho x}
\end{equation}
defines a martingale $(G^\sharp_t)_{0\leq t\leq T}$.
Then
\begin{equation}\label{dPsharp}
\frac{dP^\sharp}{dP}=\mathcal E(\tilde N^\sharp)_T
\end{equation}
defines a probability measure $P^\sharp\sim P$ on $\mathcal F_T$.
The process $(X_t)_{0\leq t\leq T}$ is a semimartingale
under $P^\sharp$; its semimartingale
characteristics with respect to the zero truncation function
are $(B^\sharp,C^\sharp,\nu^\sharp)$ which are given by
\begin{equation}\label{nu-sharp}
dB^\sharp_t=b^\sharp_tdt,\quad
dC^\sharp_t=c^\sharp_tdt,\quad
\nu^\sharp(dt,dx)=
F^\sharp(t,dx)dt,
\end{equation}
where
\begin{equation}\label{F-sharp}
b^\sharp_t=
\mu+(\beta+\theta^\sharp_t)V_{t-},\quad
c^\sharp_t=V_{t-},\quad
F^\sharp(t,dx)=Y^\sharp(t,x)\lambda U_\rho(dx).
\end{equation}
\end{theorem}
Proof: We can apply~\cite[Theorem~4.1, p.421]{KS} and conclude that the density
in~(\ref{dPsharp}) defines an equivalent local martingale measure for $e^X$.
By the Girsanov Theorem for general semimartingales,
\cite[3.24, p.172f]{JS}, the characteristics of $X$ under $P^\sharp$
follow.~\hfill\qed

In Appendix~\ref{app-a} we give sufficient conditions, that the
solution~$\theta^\sharp$ exists, which is then of the form
$\theta^\sharp_t=\phi^\sharp(V_{t-})$ for some Borel
function~$\phi^\sharp:\mathbb R_+\to\mathbb R$,
and $G^\sharp$ is a proper martingale and thus a density process.
\begin{remark}
If we used the identity truncation function we had
$b^\sharp_t=\mu+\lambda k_\rho'(\theta^\sharp_t)+(\beta+\theta^\sharp_t)V_{t-}$.
\end{remark}
\begin{remark}
From~(\ref{nu-sharp}) and~(\ref{F-sharp}), we see that in general the
third characteristic of~$X$, and thus of~$Z$,
under $P^\sharp$ will be non-deterministic and depend on time,
hence~$Z$ is not a \Levy{} process under
$P^\sharp$ any more.
\end{remark}
\begin{remark}
Analyzing the above calculations we see that the concrete dynamics of
the volatility process does not come into play, so analogous results
hold for a quite general class of stochastic volatility models with jumps,
including, for example, the Bates model~\cite{Bates}.
\end{remark}
%
\subsection{The Esscher martingale transform for linear processes}
Let us look at the Esscher martingale transform for linear processes
as described in \cite[Theorem~4.4, p.423]{KS}.
Note, that in our notation the discounted asset price $S$ satisfies
$S=S_0\mathcal E(\tilde X)$ and we must be careful not to
confuse $\tilde K^X$ and $K^{\tilde X}$.
We have to find the solution to
\begin{equation}\label{DK=0}
DK^{\tilde X}(\theta)_t=0.
\end{equation}
Suppose now we can establish a solution $\theta^*_t$
to that equation, almost surely for all $t\in[0,T]$, and
$G^*_t=e^{\theta^*\cdot\tilde X_t-K^{\tilde X}(\theta^*)_t}$
defines a martingale $(G^*_t)_{0\leq t\leq T}$, then
we can define a probability measure $P^*$ by
\begin{equation}
\frac{dP^*}{dP}=e^{\theta^*\cdot\tilde X_T-K^{\tilde X}(\theta^*)_T}.
\end{equation}
This measure is then the Esscher martingale transform for the linear
process~$\tilde X$. If there is no solution with the required properties,
we say the Esscher martingale transform for the linear process does not exist.
\begin{theorem}
Suppose there is $\theta^*\in L(\tilde X)$, such that $\theta^*\cdot\tilde X$
is exponentially special,
\begin{equation}\label{eq-star}
DK^{\tilde X}(\tilde\theta^*)_t=0,
\end{equation}
and
\begin{equation}\label{dP*}
G^*_t=\mathcal E(\tilde N^*)_t
\end{equation}
with
\begin{equation}
\tilde N_t^*=\int_0^t\psi^*_sdW_s+\int_0^t\int(Y^*(s,x)-1)(\mu_X-\nu)(dx,ds),
\end{equation}
\begin{equation}
\psi^*_t=\theta^*_t\sqrt{V_{t-}}
\end{equation}
and
\begin{equation}
Y^*(t,x)=e^{\theta^*_t(e^x-1)}
\end{equation}
defines a martingale $(G^*_t)_{0\leq t\leq T}$.
Then
\begin{equation}
\frac{dP^*}{dP}=\mathcal E(\tilde N^*)_T
\end{equation}
defines a probability measure $P^*\sim P$ on $\mathcal F_T$.
The process $(X_t)_{0\leq t\leq T}$ is a semimartingale
under $P^*$ with semimartingale
characteristics $(B^*,C^*,\nu^*)$ given by
\begin{equation}\label{nu-star}
dB^*_t=b^*_tdt,\quad
dC^*_t=c^*_tdt,\quad
\nu^*(dt,dx)=F^*(t,dx)dt,
\end{equation}
where
\begin{equation}\label{F-star}
b^*_t=
\mu+(\beta+\theta^*_t)V_{t-},\quad
c^*_t=V_{t-},\quad
F^*(t,dx)=Y^*(t,x)\lambda U_\rho(dx).
\end{equation}
\end{theorem}
We can apply~\cite[Theorem~4.4, p.423]{KS} and conclude that the density
in~(\ref{dP*}) defines an equivalent local martingale measure
for~$\mathcal E(\tilde X)$.
So we can apply again the Girsanov Theorem for general semimartingales,
\cite[3.24, p.172f]{JS}, to derive the characteristics of~$X$
under~$P^*$.~\hfill\qed

In Appendix~\ref{app-b} it is shown that there exists
always a measurable
function $\phi^*:\mathbb R_+\to\mathbb R$, such that
$\vartheta^*_t=\phi^*(V_{t-})$ is a solution to~(\ref{DK=0}),
and sufficient conditions are given that
$G^*$ is a proper martingale and thus a density process.
\begin{remark}
If we used the identity truncation function we had
$b^*_t=\mu+\lambda \tilde k_\rho'(\theta^*_t)+(\beta+\theta^*_t)V_{t-}$
for the first differential characteristic.
\end{remark}
\begin{remark}
From~(\ref{nu-star}) and~(\ref{F-star}), we see that in general the
third characteristic of~$X$, and thus of~$Z$,
under $P^*$ will be non-deterministic and depend on time,
hence~$Z$ is not a \Levy{} process under
$P^*$ any more.
\end{remark}
\begin{remark}
A comparison of \cite[Theorem~4.3, p.477]{CS2005b}
resp.~\cite[Theorem~3.3., p.8]{CS2006}
and \cite[Theorem~4.4 p.423]{KS}
indicates that, for general semimartingales, the linear Esscher martingale
transform and the minimum entropy-Hellinger martingale measure
coincide, at least under the assumption \cite[(3.5) p.8]{CS2006},
which is equivalent to the existence of all exponential moments
of $F(t,dx)$.
This assumption holds for the Poisson toy example studied
in the next section, but not for
BNS models with a BDLP with semi-heavy tails, such as the
$\Gamma$-OU and the IG-OU model.

We conjecture that under the weaker conditions granting the existence
of the linear Esscher transform given in the appendix,
the linear Esscher measure and the minimum entropy-Hellinger martingale measure
coincide also for the last two models mentioned.
As systematic and and rigorous investigation of this relationship
involves some technicalities and is left open for future research.
\end{remark}
\section{Examples}\label{Sec-Examples}
\subsection{The Poisson toy example}
\subsubsection{Exponential Esscher martingale transform}
This model is used for illustrative purposes, since all calculations are explicitly possible.
Suppose
\begin{equation}
Z_t=\delta N_t
\end{equation}
where $\delta>0$ is the jump size and $N$ is a standard Poisson process with
intensity parameter $\gamma>0$. Then
\begin{equation}
k(\theta)=\gamma(e^{\delta\theta}-1)
\end{equation}
and the solution of equation~(\ref{eq-sharp}) is
\begin{equation}
\theta^\sharp_t=
-\frac{\mu+\tilde\beta V_{t-}}{V_{t-}}
-\frac1{\rho\delta}w\left(
\frac{\delta\rho \lambda\gamma(e^{\delta\rho}-1)}{V_{t-}}\exp\left(
-\delta\rho\frac{\mu+\beta V_{t-}}{V_{t-}}
\right)
\right),
\end{equation}
where $w$ is known as (the principal branch of) the
Lambert~$W$ (or polylogarithm) function.
The function $w$ is available in Mathematica, Maple, and many other
computer packages and libraries. Basically it is the inverse function
of $xe^x$.
For further references and
code for numerical evaluation see \cite{LaW} and \cite{WAPR}.
We need to know here only, that $w$ is strictly increasing from $0$ to $\infty$
as $x$ goes from $0$ to $\infty$.

For this model we have $E[e^{\xi Z_1}]<\infty$ for all $\xi\in\mathbb R$,
so the condition~(\ref{theta1}) in Lemma~\ref{sharp-martingale} is satisfied,
and the exponential Esscher martingale transform exists.
\subsubsection{Linear Esscher martingale transform}
The jumps of $\tilde X$ are
\begin{equation}
\Delta\tilde X_t=e^{\rho\Delta X_t}-1
\end{equation}
and since we have in the Poisson toy model only one jump size, this implies, that
we can write
\begin{equation}
\tilde X_t=\tilde\delta_\rho N_t
\end{equation}
where
\begin{equation}
\tilde\delta_\rho=e^{\rho\delta}-1.
\end{equation}
So the cumulant function is of the same form we have seen in the previous section, namely
\begin{equation}
\tilde k_\rho(z)=\gamma(e^{\tilde\delta_\rho z}-1),
\end{equation}
and its derivative is
\begin{equation}
\tilde k_\rho'(z)=\gamma\tilde\delta_\rho e^{\tilde\delta_\rho z}.
\end{equation}
For the linear Esscher transform we have to solve~(\ref{eq-star}),
which becomes
\begin{equation}
\tilde b_t+\tilde c_t\theta+\lambda\gamma\tilde\delta_\rho e^{\tilde\delta_\rho \theta}=0.
\end{equation}
The solution will be given, again using the Lambert $w$ function, as
\begin{equation}
\theta^*_t=
-\frac{\mu+\tilde\beta V_{t-}}{V_{t-}}
-\frac1{\tilde\delta_\rho}w\left(
\frac{\lambda\gamma\tilde\delta_\rho^2}{V_{t-}}\exp\left(
-\tilde\delta_\rho\frac{\mu+\tilde\beta V_{t-}}{V_{t-}}
\right)
\right).
\end{equation}
As we have $E[e^{\xi Z_1}]<\infty$ for all $\xi\in\mathbb R$,
the condition~(\ref{betaZ}) in Lemma~\ref{star-martingale} is satisfied,
and the linear Esscher martingale transform exists.
\subsection{The $\Gamma$-OU example}
\subsubsection{Exponential Esscher martingale transform}
Suppose we have a stationary variance with $\Gamma(\delta,\gamma)$ distribution.
Then the BDLP is a compound Poisson process with exponential jumps and
has cumulant function
\begin{equation}
k(\theta)=\frac{\delta\theta}{\gamma-\theta}
\end{equation}
for $\Re\theta<\gamma$.
For the exponential Esscher transform we must have  $\theta^\sharp_t=\phi^\sharp(V_{t-})$
where the function $\phi^\sharp$ is obtained by solving
the equation
\begin{equation}
\mu+\tilde\beta v+v\phi
+\lambda\frac{\delta\rho(\phi+1)}{\gamma-\rho(\phi+1)}
-\lambda\frac{\delta\rho\phi}{\gamma-\rho\phi}=0.
\end{equation}
This equation can be transformed into a cubic polynomial equation
in~$\phi$ and thus, a real solution always exists.

Lemma~\ref{sharp-martingale}
provides sufficient conditions for~$G^\sharp$ to be a true martingale,
namely conditions~(\ref{theta1}), that
can be written in this case as
\begin{equation}
\rho\left[\frac{(\mu+\lambda \delta\rho/(\gamma-\rho))_+}{V_0e^{-\lambda T}}+\tilde\beta\right]_+<\gamma
\end{equation}
and
\begin{equation}\label{GammaTheta}
\frac12
\max\left\{
\left[\frac{(\mu+\lambda \delta\rho/(\gamma-\rho))_+}{V_0e^{-\lambda T}}+\tilde\beta\right]_+,
\left[\frac{-(\mu+\lambda \delta\rho/(\gamma-\rho))_-}{V_0e^{-\lambda T}}+\tilde\beta\right]_-
\right\}^2<\gamma
\end{equation}
\subsubsection{Linear Esscher martingale transform}
We have to solve the equation
\begin{equation}
\mu+\tilde\beta\theta+v\theta+
\lambda\int_0^\infty
e^{\theta(e^{\rho x}-1)}(e^{\rho x}-1)\delta x^{-1}e^{-\gamma x}dx=0.
\end{equation}
We do not have a closed form expression for the integral in the last equation,
but we know from Lemma~\ref{always}, that there is always a real solution,
that could be obtained numerically.

To apply Lemma~\ref{star-martingale} we must have~(\ref{GammaTheta})
and in that case we can conclude that the linear Esscher martingale transform
exists.
\subsection{The IG-OU example}
\subsubsection{Exponential Esscher martingale transform}
The cumulant function of the BDLP in the IG-OU model is
\begin{equation}
k(\theta)=\frac{\delta\theta}{\sqrt{\gamma^2-2\theta}}
\end{equation}
for $\Re(\theta)<\gamma^2/2$. To determine the exponential Esscher martingale
transform we have to find a solution to (\ref{KX1=0}),
which becomes equivalent to solving
$f(\theta;V_{t-})=0$ with $\theta>\gamma^2/(2\rho)$,
where
\begin{equation}
f(\theta;v)=(\mu+\tilde\beta v)+v\theta+
\lambda\delta\rho\left[
\frac{\theta+1}{\sqrt{\gamma^2-2\rho(\theta+1)}}
-
\frac{\theta}{\sqrt{\gamma^2-2\rho\theta}}
\right].
\end{equation}
In the notation of Lemma~\ref{lem-xi} we have
\begin{equation}
\xi_1=\gamma^2/2,\qquad
\ell_0=\infty
\end{equation}
and so we know there is always a solution.
The equation for $\theta^\sharp$ can be transformed into a
polynomial equation of eighth order.

The conditions~(\ref{theta1}) in Lemma~\ref{sharp-martingale}
for this model can be written as
\begin{equation}
\rho\left[\frac{(\mu+\lambda \delta\rho/\sqrt{\gamma^2-2\rho})_+}{V_0e^{-\lambda T}}+\tilde\beta\right]_+,
<\frac{\gamma^2}{2}
\end{equation}
and
\begin{equation}\label{IG2Theta}
\frac12
\max\left\{
\left[\frac{(\mu+\lambda \delta\rho/\sqrt{\gamma^2-2\rho})_+}{V_0e^{-\lambda T}}+\tilde\beta\right]_+,
\left[\frac{-(\mu+\lambda \delta\rho/\sqrt{\gamma^2-2\rho})_-}{V_0e^{-\lambda T}}+\tilde\beta\right]_-
\right\}^2\leq\frac{\gamma^2}{2}
\end{equation}
and if so, the exponential Esscher martingale transform exists.
\subsubsection{Linear Esscher martingale transform}
We have to solve the equation
\begin{equation}
\mu+\tilde\beta\theta+v\theta+
\lambda\int_0^\infty
e^{\theta(e^{\rho x}-1)}(e^{\rho x}-1)
\frac{\delta}{\sqrt{2\pi}}x^{-3/2}e^{-\gamma^2 x/2}dx=0.
\end{equation}
We do not have a closed form expression for the integral in the last equation,
but we know from Lemma~\ref{always}, that there is always a real solution,
that could be obtained numerically.

To apply Lemma~\ref{star-martingale} we must have (\ref{IG2Theta})
and in that case we can conclude that the linear Esscher martingale transform
exists.
\section{Other equivalent martingale measures for BNS models}\label{sec-other}
E.~Nicolato and E.~Venardos \cite{NV} have given a complete characterization
of all equivalent martingale measures for BNS models through the following
theorem (slightly reformulated, see Remark~\ref{reform} below).
\begin{theorem}\label{D}
Let $Q$ be an EMM for the BNS model.
Then the corresponding density process
is given by the stochastic exponential
\begin{equation}\label{NV6a}
G^Q_t=\mathcal{E}({\tilde N^Q})_t,
\end{equation}
where
\begin{equation}\label{NV6b}
\tilde N^Q_t=\int_0^t\psi^Q_sdW_s+\int_0^t\int(Y^Q(s,x)-1)(\mu_X-\nu)(dx,ds),
\end{equation}
and where $\psi^Q$ is a predictable process and $Y^Q$ is
a strictly positive predictable function such that
\begin{equation}
\int_0^T(\psi_s^Q)^2ds<\infty\qquad\mbox{$P$-a.s.}
\end{equation}
and
\begin{equation}\label{NV7}
\int_0^T\int_0^{\infty}\left(\sqrt{Y^Q(s,x)}-1\right)^2U_\rho(dx)<\infty
\qquad\mbox{$P$-a.s.}
\end{equation}
The function $Y^Q$ and the process $\psi^Q$ are related by
\begin{equation}\label{NV8}
\mu+(\beta+\frac12)V_{t-}+\sqrt{V_{t-}}\psi^Q_t
+\int_0^{\infty}Y^Q(x,t)(e^x-1)\lambda U_\rho(dx)=0
\qquad\mbox{$dP\otimes dt$-a.e.}
\end{equation}
The process $(X_t)_{0\leq t\leq T}$ is a semimartingale
under $Q$ with semimartingale
characteristics $(B^Q,C^Q,\nu^Q)$ with respect to the zero truncation
function are given by
\begin{equation}\label{nu-Q}
dB^Q_t=b^Q_tdt,\quad
dC^Q_t=c^Q_tdt,\quad
\nu^Q(dt,dx)=
F^Q(t,dx)dt,
\end{equation}
where
\begin{equation}\label{F-Q}
b^Q_t=-\frac12V_{t-}-\int(e^x-1)Y^Q(t,x)\lambda U_\rho(dx),\quad
c^Q=V_{t-},\quad
F^Q(t,dx)=Y^Q(t,x)\lambda U_\rho(dx).
\end{equation}
\end{theorem}
\begin{remark}
If we used the identity truncation function we had
$b^Q_t=-\frac12V_{t-}-\int(e^x-1-x)Y^Q(t,x)\lambda U_\rho(dx)$
for the first differential characteristic.
\end{remark}
\begin{remark}\label{reform}
The careful reader will note, that our $b^Q$ and $Y^Q$ have
a meaning slightly different from that appearing in \cite{NV},
and consequently equation~(\ref{NV8}) is modified.
The reasons are that, while Nicolato and Venardos work with the jump measure
of the process $(Z_{\lambda t})$, we use the jump measure of the process $X$ in order
to be notationally consistent with \cite{KS} and the rest of our paper,
see~(\ref{muXZ}). Moreover in \cite{NV} $b^Q$ and $Y^Q$ correspond to
the stochastic exponential, and the identity truncation function is used.
Finally \cite{NV} allow a riskless interest rate $r\geq0$, whereas we use discounted quantities
throughout the paper.
\end{remark}
\begin{remark}
We recall that the predictable process $\psi^Q$ and the predictable function
$Y^Q$ can be interpreted as the market price of risk associated respectively
to the diffusion and the jump part of the price process.
\end{remark}
\subsection{The minimal martingale measure for BNS models\label{Sec-Minimal}}
We will see in Section~\ref{simplifications} that in the BNS model without leverage
the minimal martingale measure coincides with both Esscher martingale transforms.
In this section we compute the minimal martingale measure for the BNS model with leverage,
and show in the general case the measures do not coincide.
\begin{theorem}\label{thm-flat}
Let
\begin{equation}\label{tildeNflat}
\tilde N^\flat_t=\int_0^t\psi^\flat_sdW_s+\int_0^t\int(Y^\flat(s,x)-1)(\mu_X-\nu)(dx,ds),
\end{equation}
with
\begin{equation}
\psi^\flat_t=
-\theta^\flat_t\sqrt{V_{t-}}
\end{equation}
and
\begin{equation}
Y^\flat(t,x)=
1-\theta^\flat_t(e^x-1),
\end{equation}
where
\begin{equation}
\theta^\flat_t=
\frac{\mu+\lambda\kappa(\rho)+\tilde\beta V_{t-}}{V_{t-}
+\lambda(\kappa(2\rho)-2\kappa(\rho))}.
\end{equation}
If
\begin{equation}\label{deltaflat}
\Delta\tilde N^\flat_t>-1
\end{equation}
and
\begin{equation}
G^\flat_t=\mathcal E(\tilde N^\flat)_t
\end{equation}
is a martingale, then
the minimal martingale measure~$P^\flat$ on~$\mathcal F_T$
exists as a probability measure, $P^\flat\sim P$, and
\begin{equation}
\frac{dP^\flat}{dP}=\mathcal E(\tilde N^\flat)_T.
\end{equation}
The process $(X_t)_{0\leq t\leq T}$ is a semimartingale
under $P^\flat$. Its
characteristics $(B^\flat,C^\flat,\nu^\flat)$ with respect
to the zero truncation function are given by
\begin{equation}
dB^\flat_t=b^\flat_tdt,\quad
dC^\flat_t=c^\flat_tdt,\quad
\nu^\flat(dt,dx)=F^\flat(t,dx)dt,
\end{equation}
where
\begin{equation}
b^\flat_t=\mu+\left(\beta-\theta^\flat_t\right)V_{t-},\quad
c^\flat_t=V_{t-},\quad
F^\flat(t,dx)=Y^\flat(t,x)\lambda U_\rho(dx).
\end{equation}
\end{theorem}
Proof: We have Doob-Meyer decomposition $S=S_0+A+M$
where
\begin{equation}
dA_t=(\mu+\lambda k(\rho)+\tilde\beta V_{t-})S_{t-}dt
\end{equation}
and
\begin{equation}
dM_t=\sqrt{V_{t-}}S_{t-}dW_t
+\int_0^\infty(e^x-1)S_{t-}(\mu_X-\nu)(dx,dt).
\end{equation}
The quadratic variation is
\begin{equation}
d[M]_t=S_{t-}^2V_{t-}dt
+\int_0^\infty S_{t-}^2(e^x-1)^2\mu_X(dx,dt).
\end{equation}
The predictable quadratic variation is thus
\begin{equation}
d\langle M\rangle_t=S_{t-}^2(V_{t-}+\lambda(k(2\rho)-2k(\rho)))dt.
\end{equation}
Let us define the process
\begin{equation}
\alpha_t=\frac{dA_t}{d\langle M\rangle_t},
\end{equation}
which is here
\begin{equation}
\alpha_t=\frac{\mu+\lambda k(\rho)+\tilde\beta
V_{t-}}{V_{t-}+\lambda(k(2\rho)-2k(\rho))}
S_{t-}^{-1}.
\end{equation}
Using these processes the density of the minimal martingale measure is given by
$\mathcal E(-\int\alpha dM)$, see~\cite[p.557]{SchweizerTour}.\hfill\qed

In Appendix~\ref{app-c} we give sufficient conditions, granting that the
process $G^\flat$ is positive and a proper martingale,
and thus a density process.
\begin{remark}
If we used the identity truncation function we had
$b_t^\flat=\mu+\left(\beta-\theta^\flat_t\right)V_{t-}
+(1-\theta^\flat_t)(k'(\rho)-\rho\lambda\zeta)$.
\end{remark}
\begin{remark}
We see from the above calculations that the mean-variance-tradeoff process for the BNS
model with leverage is
\begin{equation}
K_t=\int_0^t
\frac{(\mu+\lambda k(\rho)+\tilde\beta V_{s-})^2}{V_{s-}
+\lambda(k(2\rho)-2k(\rho))}ds,
\end{equation}
and so it is not deterministic.
\end{remark}
\begin{remark}
Related to the description of the minimal martingale measure
is the minimal-variance strategy
which has been explicitly calculated for European options
in the BNS model in~\cite{abel}.
\end{remark}
\subsection{Structure preserving martingale measures}
In this section we want to examine the behavior of the class of equivalent
martingale measures for the BNS models
which preserve the model structure, in order to
compare them with the measures we obtained in the previous sections.

Under an arbitrary EMM~$Q$ it could be possible,
that $Z$ is not a \Levy{} process,
that $(W^Q,Z)$ are not independent,
and thus under~$Q$ the log-price process is no longer described by a BNS model.
We need a strong characterization of the subclass of EMMs which preserve
the model structure.
This class of measures has also been characterized in \cite{NV} with
the following theorem.
\begin{theorem}\label{E}
Let $y(x)$ be a function $y:\mathbb{R}_+\rightarrow\mathbb{R}_+$
such that
\begin{equation}
\int(\sqrt{y(x)}-1)^2U_\rho(dx)<\infty.
\end{equation}
Then the process given by
\begin{equation}\label{NV9}
\psi^y_t=-V_{t-}^{-1/2}\left(\mu+\tilde\beta V_{t-}+\lambda k^y(\rho)\right)
\end{equation}
where
\begin{equation}\label{NV10}
k^y(\theta)=\int_0^\infty\left(e^{\theta x}-1\right)y(x)U(dx)
\end{equation}
for $\Re(\theta)<0$, is such that
\begin{equation}
\int_0^T\psi_s^2ds<\infty\qquad\mbox{$P$-a.s.},
\end{equation}
and
\begin{equation}\label{NV11a}
G^y_t=\mathcal{E}({\tilde N^y})_t,
\end{equation}
where
\begin{equation}\label{NV11b}
\tilde N^y_t=\int_0^t\psi^y_sdW_s+\int_0^t\int_0^\infty(y(x)-1)(\mu_X-\nu)(dx,ds),
\end{equation}
is a density process. The probability measure defined
by $dQ^y=G^y_TdP $ is an EMM on $\mathcal F_T$ for the BNS model.
The process $(X_t)_{0\leq t\leq T}$ is a semimartingale
under $Q$ with semimartingale
characteristics $(B^Q,C^Q,\nu^Q)$ given by
\begin{equation}\label{nu-y}
dB^y_t=b^y_tdt,\quad
dC^y_t=c^y_tdt,\quad
\nu^y(dt,dx)=
F^y(t,dx)dt,
\end{equation}
where
\begin{equation}\label{F-y}
b^y_t=-\frac12V_{t-}-\int(e^x-1)y(x)\lambda U_\rho(dx),\quad
c^y_t=V_{t-},\quad
F^y(t,dx)=y(x)\lambda U_\rho(dx).
\end{equation}
The process $W_t^y=W_t-\int_0^t\psi^y_sds$ is a $Q$-Brownian motion
and $Z_{\lambda t}$ is a $Q^y$-\Levy{} process, such that $Z_1$
has \Levy{} measure $U^y(dx)=y(x)U(dx)$ and
cumulant transform $k^y(\theta)$,
and the processes $W^y$ and $Z$ are $Q^y$-independent.

Conversely, for any $Q$ satisfying the requirements
above, there exists a
function $y:\mathbb{R}_+\rightarrow\mathbb{R}_+$ with
$\int\limits_{0}^{\infty }(\sqrt{y(x)}-1)^{2}U(dx)<\infty $,
such that $Q$ coincides with $Q^y$.
\end{theorem}
\begin{remark}
If we used the identity truncation function we had
$b^y_t=-\frac12V_{t-}-\int(e^x-1-x)y(x)\lambda U_\rho(dx)$
for the first differential characteristic.
\end{remark}
\begin{remark}
Structure Preserving Equivalent Martingale Measures (from now on SPEMM) are
relevant since they allow to obtain some analytical results for
option pricing, see \cite{NV}. Since the Laplace transform of
log-prices has a simple expression, using a transform-based technique the
authors can obtain some closed-form formulas for the price of European
options in several relevant cases.
\end{remark}
\begin{remark}
The structure preserving measures are in general not Esscher transforms
with respect to $X$ or $\tilde X$. In the special case, when only the law of
the BDLP is changed such that $y(x)=e^{\theta x}$, the density process is given by
\begin{equation}
L_t=e^{\theta Z_{\lambda t}-\lambda k^y(\theta)t}
\end{equation}
and thus we have an Esscher transform with respect to the \Levy{}
process $(Z_{\lambda t})$.
\end{remark}
Whenever the distribution of the BDLP belongs to the same parametric class (such as gamma or inverse Gaussian, for example)
under the original and under an equivalent martingale measure, we say the measure change is
{\em distribution preserving}. The distribution preserving measures are obviously a subclass of the structure
preserving measures.
\begin{remark}[Uniqueness]
It follows from the examples below, that neither the equivalent martingale measures,
the structure preserving or the distribution preserving martingale measures are unique
in general. But, the measure that does not change the law of the~BDLP $Z$,
is unique.
In this case we have $Y=1$, and equation (\ref{NV9}) determines
uniquely the change of drift for the Brownian motion $W$. This measure is trivially
distribution preserving.
Economically this choice of martingale measure corresponds to the (questionable) idea,
that the jumps represent only {\em non-systematic} risk that is not reflected in derivatives
prices, cf.~\cite[p.133]{Merton1976}.
\end{remark}
\subsection{The minimal entropy martingale measure\label{Sec-Entropy}}
An important equivalent martingale measure, that can be defined for a wide
class of general semimartingales is the minimal entropy martingale
measure (MEMM). This measure is also relevant for its connection
with utility maximization with respect to exponential utility.
The definition of the MEMM and a systematic investigation on this connection
is given in \cite{MF,FF}.
  For exponential \Levy{} models the MEMM coincides with the linear Esscher
martingale transform, see \cite{ES,Fumi,levyrep}.
A natural question is whether this property holds also more generally,
and in particular for BNS models.
We will see below by direct comparison, that in the non-leverage-case the
answer is negative.
The minimal entropy martingale measure for the BNS model in the leverage
case, i.e., when $\rho\neq 0$, has been obtained by T.~\Rheinlaender{}
and G.~Steiger. In~\cite[Corollary 4.5, p.1340f]{RS}
they provide a representation formula in terms of the solution of a
semi-linear integro-PDE, but
from this representation formula it seems difficult
to make a direct comparison in the general case.
In the simple concrete example of the Poisson toy model it is possible
to verify explicitly that the two measures do not coincide. This
leads to the conclusion that in general, the MEMM and the (linear)
Esscher transform for BNS models are different.

As \Rheinlaender{} and Steiger already remarked
in~\cite[Remark~4.4.4, p.1339]{RS}, the MEMM does not
preserve the independence of increments of the BDLP,
thus is not a structure preserving measure.
\section{Simplifications for the BNS model without leverage\label{simplifications}}
\subsection{The Esscher martingale transform for BNS models without leverage}
Let us now examine the simplification that occur in the simpler situation of without leverage,
i.e., when $\rho=0$.
In this case it turns out, that both the Esscher martingale transforms
for exponential and for linear processes as well as the minimal martingale measure
coincide. In fact, this is true, for all \Ito{} process, as we see from the following lemma.
\begin{lemma}
Suppose the logarithmic return process $X$ satisfies
\begin{equation}\label{dX-general}
dX_t=\mu_tdt+\sigma_tdW_t
\end{equation}
with $W$ a standard Brownian motion, and $\mu$ and $\sigma$ are adapted processes,
such that (\ref{dX-general}) is well-defined.
Then the Esscher martingale transforms for the exponential process $e^X$,
the Esscher martingale transform for the linear process $\tilde X$, and
and the minimal martingale measure either exist and coincide, or neither of them exists.
\end{lemma}
Proof: The modified Laplace cumulant process $K^X(\theta)$ of $X$ in $\theta$
is given by
\begin{equation}
K^X(\theta)_t=\int_0^t\tilde\kappa^X(\theta)_sds,
\end{equation}
where
\begin{equation}
\tilde\kappa^X(\theta)_t=
\mu_t\theta_t+\frac12\sigma_t^2\theta_t^2.
\end{equation}
Finding the parameter process $\theta$ that turns the exponential process $e^X$
into a martingale requires to solve
\begin{equation}
\tilde\kappa^X(\theta+1)-\tilde\kappa^X(\theta)=0,
\end{equation}
i.e.,
\begin{equation}
\mu_t(\theta_t+1)+\frac12\sigma_t^2(\theta_t+1)^2
\mu_t\theta_t+\frac12\sigma_t^2\theta_t^2=0.
\end{equation}
That gives the solution
\begin{equation}
\theta^\sharp_t=
-\frac{\mu_t+\frac12\sigma_t^2}{\sigma_t^2},
\end{equation}
and we obtain
\begin{equation}
\frac{dP^\sharp}{dP}=\exp\left(
\int_0^T\theta^\sharp_tdX_t-K^X(\theta^\sharp)_t
\right),
\end{equation}
and thus
\begin{equation}\label{dPsharpdP}
\frac{dP^\sharp}{dP}=\exp\left(
-\int_0^T
\frac{\mu_t+\frac12\sigma_t^2}{\sigma_t}dW_t
-\frac12\int_0^T
\frac{(\mu_t+\frac12\sigma_t^2)^2}{\sigma_t^2}dt
\right),
\end{equation}
provided that the density process is a proper martingale.
Now let us compute the Esscher martingale transform for the linear process
$\tilde X$. We have
\begin{equation}
d\tilde X_t=(\mu_t+\frac12\sigma_t^2)dt+\sigma_tdW_t.
\end{equation}
The modified Laplace cumulant process $K^{\tilde X}(\theta)$ of $\tilde X$ in
$\theta$ is given by
\begin{equation}
K^{\tilde X}(\theta)t=\int_0^t\tilde\kappa^{\tilde X}(\theta)_sds,
\end{equation}
where
\begin{equation}
\tilde\kappa^{\tilde X}(\theta)_t=
(\mu_t+\frac12\sigma_t^2)\theta_t+\frac12\sigma_t^2\theta_t^2.
\end{equation}
We need the derivative
\begin{equation}
DK^{\tilde X}(\theta)t=\int_0^tD\tilde\kappa^{\tilde X}(\theta)_sds,
\end{equation}
where
\begin{equation}
D\tilde\kappa^{\tilde X}(\theta)_t=
(\mu_t+\frac12\sigma_t^2)+\sigma_t^2\theta_t.
\end{equation}
We have to solve $DK^{\tilde X}(\theta)=0$, which has in our case the solution
\begin{equation}
\theta^*_t=
-\frac{\mu_t+\frac12\sigma_t^2}{\sigma_t^2}.
\end{equation}
Then the density for the Esscher martingale transform for the linear process $\tilde X$ is given by
\begin{equation}
\frac{dP^*}{dP}=\exp\left(
\int_0^T\theta^*_td\tilde X_t-K^{\tilde X}(\theta^*)_t,
\right)
\end{equation}
and thus
\begin{equation}\label{dPastdP}
\frac{dP^*}{dP}=\exp\left(
-\int_0^T
\frac{\mu_t+\frac12\sigma_t^2}{\sigma_t}dW_t
-\frac12\int_0^T
\frac{(\mu_t+\frac12\sigma_t^2)^2}{\sigma_t^2}dt
\right),
\end{equation}
provided that the density process is a proper martingale.
We see that the expressions (\ref{dPsharpdP}) and (\ref{dPastdP}) coincide,
and thus $P^\sharp=P^*$. By comparing our result with the expression for the density of the minimal
martingale measure, see for example \cite[(1.1), p.28]{Schweizer1999}
we see that the Esscher martingale transforms agree with the minimal martingale measure.~\qed
\begin{remark}
As it is apparent from the proof the reason by which $P^\sharp$ and $P^*$ coincide is the fact, that
\begin{equation}
\tilde X-X=K^{\tilde X}(\theta)-K^{X}(\theta)
\end{equation}
for any parameter process $\theta$.
\end{remark}
\subsection{Minimal entropy for BNS without leverage}
We want to recall in this section some results available for the minimal
entropy martingale measure in the framework of the BNS model without leverage and we
want to compare them with the measures we have obtained in order to show
that for BNS, the MEMM and the Esscher martingale transform for the linear
process in general do not coincide. F.E.~Benth and T.~Meyer-Brandis
have obtained in \cite[Proposition 5.2, p.13]{EBMB} an explicit
expression for the MEMM
in the particular case of the BNS model without leverage, i.e.,
when the coefficient $\rho=0$. The measure is obtained as the zero
risk aversion limit of the martingale measure corresponding
to the indifference price with respect to the exponential utility function.

Under some integrability conditions they have proved, that the MEMM
is given by
\begin{equation}\label{Pe0}
\frac{dP^e}{dP}=
\frac{\exp\left[-\int_0^T\frac{\mu+\tilde\beta V_{t-}}{\sqrt{V_{t-}}}dW_{t}
-\int_0^T\frac{(\mu+\tilde\beta V_{t-})^2}{V_{t-}}dt\right]}%
{E\left[\exp(-\int_0^T\frac{(\mu+\tilde\beta V_{t-})^2}{2V_{t-}}dt)\right]}.
\end{equation}
Actually, \cite{EBMB} write $V_t$ instead of $V_{t-}$
but this does not make a difference in the present case.
It is not difficult to see that this measure does not preserve the \Levy{}
property, and thus the model structure; in order to have the \Levy{} property preservation, in fact, the
measure should be of the form (\ref{NV6a}--\ref{NV6b}) in which $y(x)$ must be
deterministic and time independent.
Moreover a direct
comparison of~(\ref{Pe0}) with~(\ref{dPsharpdP}) shows that this measure does not coincide
neither with the Esscher martingale transforms nor the minimal martingale
measure. This remark allows to conclude that BNS models have a quite
different behavior in comparison with exponential \Levy{} models with
respect to these classes of measures. In the exponential \Levy{} case, in
fact, it has been proved \cite{ES} that the MEMM coincides with
the Esscher martingale transform for the linear process and that this
measure has the special property of preserving the \Levy{} structure of the
model.
\begin{remark}
In \cite{ES} the MEMM for a particular stochastic volatility
model with \Levy{} jumps has been investigated, for which it turns out that
MEMM has the same properties of \Levy{} structure preservation and it
coincides with the Esscher martingale transform for the linear process. This
analogy with the exponential \Levy{} models breaks down for more complex
models like BNS.
\end{remark}
\begin{remark}
In contrast to the exponential \Levy{} model the minimal entropy measure is for the
BNS models not time-independent.
This means, for given $0<T_1<T_2$, the minimal entropy martingale measure for horizon $T_1$ is not obtained
as restriction of the minimal entropy martingale measure for horizon $T_2$ to $\mathcal F_{T_1}$.
This was also observed for the Stein and Stein / Heston model
in \cite{Rhe2005}.
\end{remark}

%
%
%
%
%
%
%
\subsection*{Acknowledgments} We want to thank Ole Barndorff-Nielsen
for suggesting a starting point for the present investigation and for
valuable suggestions. We also thank Alberto Barchielli, Fabio Bellini, Marco Frittelli,
Stephen Lauritzen, and Martin Schweizer for useful comments.

We thank Thorsten Rheinl\"ander and Gallus Steiger for
showing and discussing their results with us.

The subject of this report has been
developed during visits of the authors
to the
Mathematics Department of the Politecnico di Milano to
the Department of Mathematical Sciences at Aarhus, respectively.
We thank both institutes for their hospitality.

A preliminary version of this article was presented at
the Annual CAF members' meeting, January 13--14, 2005,
at Sandbjerg Gods, Denmark. We thank the participants and especially
the discussant S\o{}ren Willemann for several helpful suggestions.

\nocite{CS2005}

\appendix

\section{On existence and integrability conditions for
the exponential Esscher martingale transforms\label{app-a}}
\begin{lemma}\label{lem-xi}
Let
\begin{equation}
\xi_1=\sup\{\xi\geq0:E[e^{\xi Z_1}]<\infty\}.
\end{equation}
and
\begin{equation}
\ell_0=\inf_{\theta>\xi_1/\rho}\left[k(\rho(\theta+1))-k(\rho\theta)\right].
\end{equation}
If one of the four conditions
\begin{enumerate}
\item $\xi_1=+\infty$, or
\item $\xi_1<+\infty$ and $\ell_0=-\infty$,
\item $\xi_1<+\infty$ and $\ell_0>-\infty$, $\beta+1/2+\xi_1/\rho=0$, and $\mu+\lambda\ell_0\leq0$, or
\item $\xi_1<+\infty$ and $\ell_0>-\infty$, $\beta+1/2+\xi_1/\rho<0$, and $\displaystyle V_0e^{-\lambda T}\geq-\frac{\mu+\lambda\ell_0}{\beta+1/2+\xi_1/\rho}$,
\end{enumerate}
holds, then there is a measurable
function $\phi:\mathbb R_+\to\mathbb R$,
such that $\vartheta^\sharp_t=\phi(V_{t-})$
is a solution to~(\ref{KX1=0}).
\end{lemma}
Proof: Let
The function $k(\xi)$ is well-defined and analytic for $\xi<\xi_1$.
Let us study now the behavior of
\begin{equation}
\ell(\theta)=k(\rho(\theta+1))-k(\rho\theta).
\end{equation}
This function is well-defined for $\theta>\theta_0$, where
\begin{equation}
\theta_0=\frac{\xi_1}{\rho},
\end{equation}
and we have
\begin{equation}\label{ell-int}
\ell(\theta)=\int_0^\infty e^{\theta \rho x}(e^{\rho x}-1)U(dx).
\end{equation}
This is a Laplace transform and we can differentiate under the integral to obtain
\begin{equation}\label{ell'-int}
\ell'(\theta)=\int_0^\infty e^{\theta \rho x}(\rho x)(e^{\rho x}-1)U(dx).
\end{equation}
From~(\ref{ell-int}) we infer that $\ell(\theta)<0$, and, by monotone convergence, that
\begin{equation}
\lim_{\theta\to+\infty}\ell(\theta)=0.
\end{equation}
From~(\ref{ell'-int}) we see that $\ell(\theta)$ is increasing.
We have to solve $f(\theta,v)=0$, where
\begin{equation}\label{f}
f(\theta,v)=\mu+(\beta+\textfrac12)v+v\theta+\lambda\ell(\theta)
\end{equation}
for all $v>0$.
Under the conditions 1.\ and 2.\ we have
\begin{equation}
\inf_{\theta>\theta_0}f(\theta,v)=-\infty.
\end{equation}
Under condition 3.\ and 4.\ we have
\begin{equation}
\inf_{\theta>\theta_0}f(\theta,v)=\mu+(\beta+1/2)v+v\theta_0+\lambda\ell_0
\end{equation}
Under conditions~3.\ and~4.\ this is less or equal to zero for all $v>0$,
and thus we have a solution.\hfill\qed
\begin{lemma}\label{sharp-martingale}
Suppose $\theta^\sharp$ is a solution to~(\ref{KX1=0}). Let
\begin{equation}
N^\sharp_t=\int_0^t\theta^\sharp_sdX_s-K^X(\theta^\sharp)_t,
\end{equation}
and
\begin{equation}\label{GsharpN}
G^\sharp_t=e^{N^\sharp_t}.
\end{equation}
If
\begin{equation}\label{theta1}
E[Z_1e^{\rho\Theta^\sharp_0Z_1}]<\infty,\qquad
E[e^{\frac12(\Theta^\sharp_1)^2Z_1}]<\infty
\end{equation}
where
\begin{equation}
\Theta^\sharp_0=-\left[\frac{(\mu+\lambda k(\rho))_+}{V_0e^{-\lambda T}}+\tilde\beta\right]_+,
\qquad
\Theta^\sharp_1=
\max\left\{
\left[\frac{(\mu+\lambda k(\rho))_+}{V_0e^{-\lambda T}}+\tilde\beta\right]_+,
\left[\frac{-(\mu+\lambda k(\rho))_-}{V_0e^{-\lambda T}}+\tilde\beta\right]_-
\right\}
\end{equation}
then $(G^\sharp_t)_{0\leq t\leq T}$ is a martingale.
\end{lemma}
Proof: The equation $f(\theta,v)=0$ implies that
$\theta^\sharp_t=\phi^\sharp(V_{t-})$ is bounded.
Let us provide now concrete bounds for $\theta^\sharp$:
We have
\begin{equation}\label{thetasharpeq}
\theta^\sharp_t=-\left[\frac{\mu+\lambda\ell(\theta^\sharp_t)}{V_{t-}}+\tilde\beta\right].
\end{equation}
We have already observed that $\ell(\theta)$ is negative and increasing in $\theta$, and $V_{t-}>V_0e^{-\lambda T}$.
Distinguishing the cases $\theta^\sharp_t\leq0$ and $\theta^\sharp_t>0$ we obtain
\begin{equation}
-\left[\frac{(\mu+\lambda\ell(0))_+}{V_0e^{-\lambda T}}+\tilde\beta\right]\leq
\theta^\sharp_t
\leq0
\end{equation}
respectively
\begin{equation}
0\leq\theta^\sharp_t\leq
-\left[-\frac{(\mu+\lambda\ell(0))_-}{V_0e^{-\lambda T}}+\tilde\beta\right]
\end{equation}
The subscripts plus and minus in those inequalities denote the positive
and negative part. We note, that $\ell(0)=k(\rho)$.
Let us next consider the process $G^\sharp$. We can rewrite~(\ref{GsharpN}) as
\begin{equation}
G^\sharp_t=\mathcal E(\tilde N^\sharp)_t,
\end{equation}
where $\tilde N^\sharp$ is the exponential transform of the process $N^\sharp$.
Since $K^{X}(\theta)$ is continuous we have
\begin{equation}
\Delta N^\sharp_t=\theta^\sharp_t\rho\Delta Z_{\lambda t}
\end{equation}
and thus
\begin{equation}\label{DeltatildeNsharp}
\Delta\tilde N^\sharp_t=e^{\theta^\sharp_t\rho\Delta Z_{\lambda t}}-1.
\end{equation}
To prove the lemma we use the integrability condition
from~\cite[Theorem~III.1, p.185f]{LepingleMemin1978}.
We have to show, that
\begin{equation}
A_t=\textfrac12\langle\tilde N^{\sharp c}\rangle_t
+\sum_{s\leq t}(1+\Delta\tilde N^\sharp_s)\log(1+\Delta\tilde N^\sharp_s)-\Delta\tilde N^\sharp_s
\end{equation}
admits a  predictable compensator $B$ such that $E[e^{B_T}]<\infty$.
Using~(\ref{DeltatildeNsharp}) we obtain, that $A$ admits indeed
a predictable compensator, which is given by
\begin{equation}\label{Bsharpint}
B_t=\textfrac12\int_0^t(\theta^\sharp_s)^2V_{s-}ds
+\int_0^t\int_0^\infty
\left[
e^{\theta^\sharp_s\rho x}\theta^\sharp_s\rho x-\left(e^{\theta^\sharp_s\rho x}-1\right)
\right]U(dx)\lambda ds.
\end{equation}
Using the boundedness for $\theta^\sharp$, a Taylor expansion at $x=0$,
and the first integrability condition in~(\ref{theta1}) we see
that the second integral in~(\ref{Bsharpint}) exists, and is, in fact, uniformly bounded by a constant.
So we have~$E[e^{B_T}]<\infty$ if
\begin{equation}
E\left[\exp\left(\frac12\int_0^T(\theta^\sharp_s)^2V_{s-}ds\right)\right]<\infty.
\end{equation}
Using the bounds on $\theta^\sharp$ from above, the last
inequality is implied by
\begin{equation}\label{betaintVexp}
E\left[\exp\left(\frac12(\Theta^\sharp_1)^2\int_0^TV_{s-}ds\right)\right]<\infty.
\end{equation}
Finally using the inequality
\begin{equation}
\int_0^TV_{s-}ds\leq V_0+Z_{\lambda T}
\end{equation}
we see, that a sufficient condition for~(\ref{betaintVexp}) is the second
integrability condition in~(\ref{theta1}).\hfill\qed
\section{On existence and integrability conditions for
the linear Esscher martingale transforms\label{app-b}}
\begin{lemma}\label{always}
There exists always a measurable
function $\phi^*:\mathbb R_+\to\mathbb R$, such that
$\vartheta^*_t=\phi^*(V_{t-})$ is a solution to~(\ref{DK=0}).
\end{lemma}
Proof:
Let us study the behavior of $\tilde\ell(z)=\tilde k_\rho'(z)$: First we observe that
$\tilde M$, which was given in~(\ref{tildeM}), has bounded jumps and
so $\tilde k_\rho(z)$ and $\tilde k_\rho'(z)$
exist and are entire functions, thus in particular
continuous on~$\mathbb R$. Differentiating~(\ref{ktilderho}) yields
\begin{equation}
\tilde\ell(z)=\int_0^\infty e^{z(e^{\rho x}-1)}(e^{\rho x}-1)U(dx).
\end{equation}
Using the integrability properties of $U(dx)$ we get by dominated convergence,
\begin{equation}
\lim_{z\to+\infty}\tilde\ell(z)=0.
\end{equation}
Let consider $b>a>0$ such that $U([a,b])>0$. Then we have for $z<0$
\begin{equation}
\tilde\ell(z)\leq e^{z(e^{\rho a}-1)}(e^{\rho a}-1)U([a,b])
\end{equation}
and
\begin{equation}
\lim_{z\to-\infty}\tilde\ell(z)=-\infty.
\end{equation}
We have to solve $\tilde f(\theta,v)=0$, where
\begin{equation}
\tilde f(\theta,v)=\mu+\tilde\beta v+v\theta+\lambda\tilde\ell(\theta).
\end{equation}
As $\tilde\ell(z)$ increases from $-\infty$ at $z\to-\infty$ to zero as $z\to+\infty$,
there is a (unique) real zero of~(\ref{DK=0}) for every $v>0$.\hfill~\qed
\begin{lemma}\label{star-martingale}
Let
\begin{equation}
N^*_t=\int_0^t\theta^*_sd\tilde X_s-K^{\tilde X}(\theta^*)_t,
\end{equation}
and
\begin{equation}\label{GstarN}
G^*_t=e^{N^*_t}
\end{equation}
where $\theta^*$ is as above.
If
\begin{equation}\label{betaZ}
E[e^{\frac12(\Theta_1^*)^2 Z_1}]<\infty
\end{equation}
with
\begin{equation}
\Theta^*_1=
\max\left\{
\left[\frac{(\mu+\lambda k(\rho))_+}{V_0e^{-\lambda T}}+\tilde\beta\right]_+,
\left[\frac{-(\mu+\lambda k(\rho))_-}{V_0e^{-\lambda T}}+\tilde\beta\right]_-
\right\}
\end{equation}
then $(G^*_t)_{t\in[0,T]}$ is a martingale.
\end{lemma}
Proof:
The equation $\tilde f(\theta,v)=0$ implies that
$\theta^*_t=\phi^*(V_{t-})$ is bounded.
Let us provide now concrete bounds for $\theta^*$:
We have
\begin{equation}\label{thetastareq}
\theta^*_t=-\left[\frac{\mu+\lambda\tilde\ell(\theta^*_t)}{V_{t-}}+\tilde\beta\right].
\end{equation}
Let us observe that $\tilde\ell(\theta)$ is negative and increasing in $\theta$, and $V_{t-}>V_0e^{-\lambda T}$.
Distinguishing the cases $\theta^*_t\leq0$ and $\theta^*_t>0$ we obtain
\begin{equation}
-\left[\frac{(\mu+\lambda\tilde\ell(0))_+}{V_0e^{-\lambda T}}+\tilde\beta\right]\leq
\theta^*_t
\leq0
\end{equation}
respectively
\begin{equation}
0\leq\theta^*_t\leq
-\left[-\frac{(\mu+\lambda\tilde\ell(0))_-}{V_0e^{-\lambda T}}+\tilde\beta\right]
\end{equation}
This can be summarized by $|\theta_t^*|\leq\Theta_1^*$.
We note that $\tilde\ell(0)=k(\rho)$.
Let us next consider the process $G^*$. We can rewrite~(\ref{GstarN}) as
\begin{equation}
G^*_t=\mathcal E(\tilde N^*)_t,
\end{equation}
where $\tilde N^*$ is the exponential transform of the process $N^*$.
Since $K^{\tilde X}(\theta)$ is continuous we have
\begin{equation}
\Delta N^*_t=\theta^*_t(e^{\rho\Delta Z_{\lambda t}}-1)
\end{equation}
and thus
\begin{equation}\label{DeltatildeN*}
\Delta\tilde N^*_t=\exp\left(\theta^*_t(e^{\rho\Delta Z_{\lambda t}}-1)\right)-1.
\end{equation}
To prove the lemma we use the integrability condition from \cite{LepingleMemin1978}.
We have to show, that
\begin{equation}
A_t=\textfrac12\langle\tilde N^{\ast c}\rangle_t
+\sum_{s\leq t}(1+\Delta\tilde N^*_s)\log(1+\Delta\tilde N^*_s)-\Delta\tilde N^*_s
\end{equation}
admits a  predictable compensator $B$ such that $E[e^{B_T}]<\infty$.
Using~(\ref{DeltatildeN*}) we obtain, that $A$ admits indeed a predictable compensator, which is given by
\begin{equation}\label{Bint}
B_t=\textfrac12\int_0^t(\theta^*_s)^2V_{s-}ds
+\int_0^t\int_0^\infty
\left[
e^{\theta^*_s(e^{\rho x}-1)}\theta^*_s(e^{\rho x}-1)-\left(e^{\theta^*_s(e^{\rho x}-1)}-1\right)
\right]\lambda U(dx)ds.
\end{equation}
Using the boundedness for $\theta^*$ and a Taylor expansion at $x=0$ we see
that the second integral in~(\ref{Bint}) is bounded by a constant.
So we have~$E[e^{B_T}]<\infty$ iff
\begin{equation}
E\left[\exp\left(\frac12\int_0^T(\theta^*_s)^2V_{s-}ds\right)\right]<\infty.
\end{equation}
Using the bounds on $\theta^*$ from above, the last
inequality is implied by
\begin{equation}\label{betaintV}
E\left[\exp\left(\frac12(\Theta_1^*)^2\int_0^TV_{s-}ds\right)\right]<\infty.
\end{equation}
Finally using the inequality
\begin{equation}
\int_0^TV_{s-}ds\leq V_0+Z_{\lambda T}
\end{equation}
we see, that a sufficient condition for~(\ref{betaintV}) is~(\ref{betaZ}).\hfill\qed
\begin{remark}\label{SameTheta}
We see that $\Theta^\sharp_1=\Theta^*_1$ since $\ell(0)=\tilde\ell(0)=k(\rho)$.
This is related to the observation, that one can use the same bound for the continuous
quadratic variation parts in both the exponential and the linear Esscher martingale transforms.
\end{remark}
\section{On existence and integrability conditions for
the minimal martingale measure\label{app-c}}
\begin{lemma}
A sufficient condition for (\ref{deltaflat}) on $0\leq t\leq T$ is
\begin{equation}
\rho\leq0,\quad
(\beta+3/2)V_0e^{-\lambda T}
\geq-\mu+\lambda k(\rho)-\lambda k(2\rho),\quad
\beta\geq-\frac32.
\end{equation}
When the jumps of $Z$ are unbounded, this is also necessary.
\end{lemma}
Proof: From~(\ref{tildeNflat}) we see, that
\begin{equation}
\Delta\tilde N^\flat_t=
-\theta^\flat_t
(e^{\rho\Delta Z_{\lambda t}}-1).
\end{equation}
As $\mathbb V[e^{\rho Z_1}]=e^{k(2\rho)}-e^{2k(\rho)}>0$
we have
\begin{equation}
k(2\rho)-2k(\rho)>0.
\end{equation}
Suppose $\Delta Z_{\lambda t}>0$. Then we have
\begin{equation}
\theta^\flat_t\geq-1>-\frac1{1-e^{\rho\Delta Z_{\lambda t}}}.
\end{equation}
By the assumptions of the lemma, this inequality holds, and thus
$\Delta\tilde N^\flat_t>-1$ and $\mathcal E(\tilde N^\flat)>0$.~\hfill\qed
\begin{lemma}
If
\begin{equation}
E[e^{\frac12K_0^2Z_1}]<\infty,
\end{equation}
where
\begin{equation}
K_0=\max\left(\left|\beta+\frac12\right|,\left|\frac{\mu+\lambda
k(\rho)}{\lambda(k(2\rho)-2k(\rho))}\right|\right)
\end{equation}
then $G^\flat$ is a martingale.
\end{lemma}
Proof:
We use \cite[Theorem~III.1]{LepingleMemin1978}
to show, that $\mathcal E(\tilde N^\flat)>0$ is a proper martingale.
To apply that theorem, we consider
\begin{equation}
A_t=\textfrac12\langle\tilde N^{\flat c}\rangle_t
+\sum_{s\leq t}(1+\Delta\tilde N^\flat_s)\log(1+\Delta\tilde N^\flat_s)-\Delta\tilde N^\flat_s
\end{equation}
and show it admits the predictable compensator $B$ with $E[e^{B_T}]<\infty$.
Let us observe that
\begin{equation}
\theta^\flat_t=\frac{a+bV_{t-}}{c+V_{t-}},
\end{equation}
with the constants $a=\mu+\lambda k(\rho)$, $b=\beta+1/2$, $c=\lambda(k(2\rho)-2k(\rho))>0$.
Looking at the rational function $v\mapsto(a+bv)/(c+v)$ we get the bound
\begin{equation}\label{theta-ineq}
|\theta^\flat_t|\leq K_0.
\end{equation}
Now we have the inequality
\begin{equation}
0<1-\theta^\flat_t(e^{\rho x}-1)<1+K_0.
\end{equation}
The predictable quadratic variation of $(\tilde N^\flat)^c$ is
\begin{equation}
\langle(\tilde N^\flat)^c\rangle_t=
\int_0^t\theta_s^{\flat2}V_{s-}ds,
\end{equation}
and we have the bound
\begin{equation}
\langle(\tilde N^\flat)^c\rangle_t\leq K_0^2\int_0^tV_{s-}ds.
\end{equation}
As $\theta^\flat_t$ is bounded we have for $f(x)=x\log x+1-x$ that
$f\left(1-\theta^\flat_t(e^{\rho x}-1)\right)=\mathcal O(x^2)$
as $x\to0$ and $f\left(1-\theta^\flat_t(e^{\rho x}-1)\right)=\mathcal O(1)$ as $x\to\infty$
and consequently there is a constant $K_1>0$ such that
\begin{equation}
0\leq\int_0^\infty f\left(1-\theta^\flat_t(e^{\rho x}-1)\right)\lambda U(dx)\leq K_1.
\end{equation}
This implies that
\begin{equation}
B_t=
\int_0^t\left[
\frac12\theta_s^{\flat2}V_{s-}+
\int f\left(1-\theta^{\flat}_t(e^{\rho x}-1)\right)\lambda U(dx)\right]ds.
\end{equation}
is well-defined, bounded, and we can conclude,
that~$B$ is the predictable compensator of~$A$.
The inequalities~(\ref{theta-ineq}) and
\begin{equation}
\int_0^tV_{s-}ds\leq V_0+Z_{\lambda t}
\end{equation}
imply there is a constant $K_2>0$ such that
\begin{equation}
B_T\leq\frac12K_0^2Z_{\lambda T}+K_2.
\end{equation}
Since $Z$ is a \Levy{} process we have
$E[\exp(\frac12K_0^2Z_{\lambda T})]<\infty$ iff $E[\exp(\frac12K_0^2Z_1)]<\infty$.~\qed

%

\newcommand{\etalchar}[1]{$^{#1}$}

\end{document}